\documentclass[submission,creativecommons]{eptcs}
\providecommand{\event}{AISOS 2013} % Name of the event you are submitting to
\usepackage{amssymb}
\usepackage{breakurl}             % Not needed if you use pdflatex only.

\usepackage{graphicx}

\input{tcilatex}

\def\titlerunning{A coordination model for ultra-large scale systems of systems}
\def\authorrunning{M. L. Bujorianu, M. C. Bujorianu}
\begin{document}

\title{A coordination model for ultra-large scale systems of systems}
\author{Manuela L. Bujorianu$^{1}$ and Marius C. Bujorianu$^{2}$ \\
%EndAName
\institute{ }{\tiny .}\\
$^{1}$Institute of Mathematics, University of Warwick, UK\\
$^{2}$School of Computer Science, University of Birmingham, UK}
\maketitle

\begin{abstract}
The ultra large multi-agent systems are becoming increasingly popular due to
quick decay of the individual production costs and the potential of speeding
up the solving of complex problems. Examples include nano-robots, or systems
of nano-satellites for dangerous meteorite detection, or cultures of stem
cells for organ regeneration or nerve repair. The topics associated with
these systems are usually dealt within the theories of intelligent swarms or
biologically inspired computation systems. Stochastic models play an
important role and they are based on various formulations of the mechanical
statistics. In these cases, the main assumption is that the swarm elements
have a simple behaviour and that some average properties can be deduced for
the entire swarm. In contrast, complex systems in areas like aeronautics are
formed by elements with sophisticated behaviour, which are even autonomous.
In situations like this, a new approach to swarm coordination is necessary.
We present a stochastic model where the swarm elements are communicating
autonomous systems, the coordination is separated from the component
autonomous activity and the entire swarm can be abstracted away as a
piecewise deterministic Markov process, which constitutes one of the most
popular model in stochastic control.

\end{abstract}

\section{Introduction}

The ultra large scale systems (ULSS) represent a cross-disciplinary concept
that refers to software intensive systems with unprecedented amount of
resources and characteristics. The term was defined by Northrop and others
in \cite{Northrop06} to describe the challenges facing the US Department of
Defence.

The systems of systems (SoS) denote complex systems where each part is a
system in itself. This area is still evolving, but it is widely accepted
that parts of an SoS are systems of systems themselves with some degree of
autonomy. Their integration forms a system with more functionality and
performance that is more than simply adding the constituent behaviours.
Being complex systems, the component systems interact and emergent
behaviours appear.

In this paper, we consider aspects of systems of systems at ultra large
scale. Specifically, we focus on mathematical modelling and coordination of
such systems. We define a simple and abstract model for a component system
that we call agent. Each agent can perform a specific activity, which is
abstracted away in our model. Then we consider the case of a large number of
communicating agents. The main contribution of the paper is to describe the
evolution of the SoS as a special Markov process.

The main characteristics of systems studied in this paper are:

\begin{description}
\item  - a large, but finite, number of agents;

\item  - elements that can communicate and collaborate;

\item  - agents that are capable of complex behaviour and can be autonomous;

\item  - capability to switch between different operating regimes when they
are thought of as systems of systems.
\end{description}

\medskip Our approach addresses these issues by employing the following key
aspects:

\begin{description}
\item  - using hybrid discrete continuous models instead of using the
mathematical theory of oscillators;

\item  - the agent autonomy is modelled via hybrid automata;

\item  - the swarm coordination is treated independently from the autonomous
behaviour of each agent;

\item  - the analytic tools of statistical mechanics can be used for
studying system properties without constraining the models.;

\item  - the system of systems modelling: the individual agent dynamics and
the overall aggregate system are characterized at different scales;

\item  - availability of stochastic control techniques by abstracting away
the system of systems as a piecewise deterministic Markov process \cite{DA93}%
, which is a well-established model in control engineering.
\end{description}

\section{Problem Formulation}

Ultra large scale systems of systems (ULSoS) are composed of a very large
number of agents that can interact and coordinate to each other. An agent is
understood here as a system that uses a fixed set of rules based on
interaction with other agents and information regarding the environment in
order to change its internal state and achieve its design objective.

Understanding the causal relation between individual agent characteristics
and the collective behaviour represents the major research challenge when
dealing with ultra large scale systems of systems. Mathematical modelling
and analysis can be used for studying their aggregate dynamics, ergodic
behaviour, metastable states, causal relations between individual agents,
collective behaviour, and so on.

There exist two fundamentally different approaches for modelling systems of
interacting agents. If the number of agents is large, then a continuum
population level approach is needed, which will provide some partial
differential equations (PDE) for spatially distributed agent densities. The
models used are called \emph{Eulerian models}, and they regard the \emph{%
macroscopic level} (collective behavior) of SoS. The second approach regards
the \emph{microscopic level}, and is based on modelling of the given SoS as
a system of interactive particles (individual agents). Each particle has its
own dynamics and it is subject to specific forces of interaction coming from
the other agents. The models used are called \emph{Lagrangian models}, and
designing such models might have tremendous consequences for getting desired
collective behaviours.

Major challenges in developing analytical frameworks include nonlinearities
in the interactions, high dimensionality of the state space, possible
randomness due to the environment influences.

In this paper we construct a mathematical framework for the analysis of
ULSoS. At the microscopic level, we propose to use stochastic hybrid models
to describe the agent dynamics. The interaction of an agent with other
agents will be defined via some inputs that will modify the continuous
dynamics of the underlying agent. At the macroscopic level, we study a ULSoS
as a system of interacting stochastic hybrid systems. For the analysis
purposes, such a system of systems needs an appropriate abstraction that is
easy to handle and provides also useful insights in the dynamics structure
of the given ULSoS. This mathematical framework constitutes an initial basis
for developing formal methods for ULSoS \cite{Hinchey1,Wiels,Klaus1}.

\section{\protect\medskip Stochastic Hybrid Models}

In this section, we present two modelling paradigms for stochastic hybrid
systems. The first one is of non-diffusion type and is represented by a
class of Markov Processes called Piecewise Deterministic Markov Processes 
\cite{DA93}. The second one is of diffusion type, and it is represented by
the most general class of stochastic hybrid processes \cite{BLchapter}. In
fact, the second one is obtained by replacing the continuous deterministic
dynamical systems that appear in the description of the first one by
diffusion processes. Intuitively, the specifications of the two models are
quite similar, but the mathematical apparatus for studying the second class
is heavily based on Ito stochastic differential equations.

For the presentation of these models as hybrid automata and comparison with
other existing models, the reader is referred to \cite{Pola2003}.

\subsection{Non-Diffusion Models\label{sect_PDMP}}

The most general non-diffusion models for stochastic hybrid systems are
represented by Piecewise Deterministic Markov Processes (PDMP) \cite{DA93}).
PDMPs are examples of stochastic hybrid processes with deterministic
continuous dynamics in the operation modes. A PDMP is a Markov process $%
(x_{t})$ with two components $(q_{t},y_{t})$, where $q_{t}$ takes values in
a discrete set $Q$ and given $q_{t}=q\in Q$, $y_{t}$ takes values in an open
set $X_{q}\subset \Bbb{R}^{\mathbf{d}(q)}$ for some function $\mathbf{d}%
:Q\rightarrow \Bbb{N}$. \noindent The state space of $(x_{t})$ is equal to 
\[
\mathbf{X}=\{(q,y)|q\in Q,y\in X_{q}\}.
\]

The Borel $\sigma $-algebra of $\mathbf{X}$, denoted by $\mathcal{B}(\mathbf{%
X)}$, is the $\sigma $-algebra generated by the open sets. By convention,
when referring to sets or functions, ``measurable'' means ``Borel
measurable''. Let $\mathcal{P}(\mathbf{X})$ be the space of probability
measure on the measurable space $(\mathbf{X,}\mathcal{B})$ equipped with the
topology of weak convergence. \noindent If $\mathbf{X}$ and $\mathbf{U}$ are
nonempty topological spaces, a stochastic kernel on $\mathbf{X}$ given $U$
is a function $R(\cdot ,\cdot )$, $R:\mathbf{U\times \mathcal{B}}(\mathbf{X}%
)\rightarrow [0,1]$, or $R:\mathbf{U\rightarrow }\mathcal{P}(\mathbf{X)},$
\noindent such that $R(u,\cdot )$ is a probability measure on $\mathbf{X}$
for each fixed $u\in \mathbf{U}$, and $R(\cdot ,B)$ is a measurable function
on $\mathbf{U}$ for each fixed $B\in \mathcal{B}(\mathbf{X)}$. \noindent If $%
\mathbf{X}$ and $\mathbf{U}$ coincide, $R$ is called stochastic kernel on $%
\mathbf{X}$.

\smallskip In the remainder of this section, we briefly present the
realization of a PDMP. Assume that for each point $z=(q,y)\in \mathbf{X}$,
there exists a unique, deterministic flow $\phi _{q}(y,t)\subset X_{q}$,
determined by a differential operator $\mathcal{X}_{q}$ on $\Bbb{R}^{d(q)}$.

If for some $t_{0}\in \Bbb{R}_{+}$, $z_{0}=(q_{0},y_{0})\in \mathbf{X}$,
then $y_{t}$, where $t\geq t_{0}$ follows $\phi _{q_{0}}(y_{0},t)$ until
either $t=T_{1}$ some random time with hazard rate $\lambda $ or until $%
y_{t}\in \partial X_{q_{0}}$ (the boundary of $X_{q_{0}}$). In both cases,
the process $x_{t}$ jumps, according to a probabilistic distribution
described by a stochastic kernel $R$ to another location of the state space, 
$(q_{1},y_{1})\in \mathbf{X}$. Again, $y_{t}$ follows a deterministic flow $%
\phi _{q_{1}}(y_{1},t)$ until a random time $T_{2}$ (independent of $T_{1}$%
), or until $y_{t}\in \partial X_{q_{1}}$, etc. The jump times $T_{i}$ are
assumed to satisfy the following condition: $\Bbb{E}(\sum_{i}\mathbf{1}%
_{T_{i}\leq t})<\infty $.

A PDMP is fully described by means of three local characteristics: (i) a
global flow $\phi (y_{0},t)$ which is the solution of the following ordinary
differential equation: 
\[
\stackrel{.}{\phi }_{q}(y_{k},t)=b_{q}(\phi (y_{k},t))\text{, }t\geq T_{k}%
\text{; }\phi _{q}(y_{k},T_{k})=y_{k}. 
\]
(ii) a jump rate $\lambda :\mathbf{X}\rightarrow \Bbb{R}_{+}$; (iii) a
stochastic kernel $R:\overline{\mathbf{X}}\times \mathcal{B}(\mathbf{X}%
)\rightarrow [0,1]$.

Let $y_{t}(x)$ be the sample path of the PDMP with the start point $x$ and $%
\{(Y_{n},T_{n})|n=1,2,...\}$ be the sequence of jump times and corresponding
post-jump locations. Between two jumps the evolution represents a
deterministic dynamical system given by the flow $\phi $ and starting with $%
Y_{n}$ at time $T_{n}$, i.e. 
\[
y_{t}(x)=\phi (Y_{n},T_{n},t)\text{, }t\in [T_{n},T_{n+1})\text{.}
\]
The post-jump locations have the following probability distributions, for
any measurable set $E\in \mathcal{B}(\mathbf{X})$%
\[
\Bbb{P}_{x}(Y_{n+1}\in E|T_{1},Y_{1},...,T_{n},Y_{n},T_{n+1})=R(\phi
(Y_{n},T_{n},T_{n+1}),E)
\]
and the sojourn time in a location (or the time interval between two jumps $%
S_{n}=T_{n+1}-T_{n}$) is given by the following distribution 
\[
\Bbb{P}_{x}(S_{n}\geq t|T_{1},Y_{1},...,T_{n},Y_{n})=\exp
\{-\int_{T_{n}}^{t+T_{n}}\lambda (y(x,s))ds\}\text{.}
\]
The resulting process is a Borel right process \cite{DA93}, i.e., a
particular strong Markov process with some additional properties regarding
the state space and the continuity of the trajectories.

\subsection{Diffusion Models}

For the purposes of this paper, in the following we give a simplified
version of the general model of stochastic hybrid systems presented in \cite
{BLchapter}. A stochastic hybrid system (SHS) is a Markov process $(x_{t})$
with two components $(q_{t},z_{t})$, where $q_{t}$ takes values in a
discrete set $Q$ and given $q_{t}=q\in Q$, $y_{t}$ takes values in an open
set $X_{q}\subset \Bbb{R}^{\mathbf{d}(q)}$ for some function $\mathbf{d}%
:Q\rightarrow \Bbb{N}$. \noindent The state space of $(x_{t})$ is equal to $%
\mathbf{X}=\{(q,y)|q\in Q$, $y\in X_{q}\}$. Usually, the state space $%
\mathbf{X}$ is embedded in an Euclidean space $\Bbb{R}^{n}$. The closure $%
\overline{\mathbf{X}}$ can be partitioned into a boundary $\mathbf{X}%
_{\delta }$ and interior $\mathbf{X}_{o}$, that will play an important role
in defining the hybrid behaviour. The boundary $\mathbf{X}_{\delta }$ will
play the role of guards from the classical hybrid automata modelling.

Under standard assumptions an SHS can be uniquely characterized by: (i) a
vector field: $b:\mathbf{X}\rightarrow \Bbb{R}^{d}$, (ii) a matrix: $\sigma :%
\mathbf{X}\rightarrow \Bbb{R}^{d\times m}$ that is a $\Bbb{R}^{d}$-valued
matrix, $m\in \Bbb{N}$, (iii) an intensity function or jump rate: $\lambda :%
\mathbf{X}\rightarrow \Bbb{R}_{+}$, and (iv) stochastic kernels: $R_{o}:%
\mathbf{X}_{o}\rightarrow \mathcal{P}(\mathbf{X)}$, and $R_{\delta }:\mathbf{%
X}_{\delta }\rightarrow \mathcal{P}(\mathbf{X)}.$

\textbf{\ }In each mode $X^{q}$, the continuous evolution is driven by the
following stochastic differential equation (SDE) 
\begin{equation}
dz_{t}^{q}=b(q,z_{t}^{q})dt+\sigma (q,z_{t}^{q})dW_{t},  \label{SDE_mode}
\end{equation}
where $(W_{t},t\geq 0)$ is the $m$-dimensional standard Wiener process in a
complete probability space. The discrete component remains constant, i.e., $%
q_{t}=q$.

In the interior of the state space $\mathbf{X}_{o}$, the process may have
discrete transitions with the rate $\lambda (x)$ when the process is at
state $x$, independently of the process history. Then the process is
transferred immediately to a new state randomly according to the stochastic
kernel $R_{o}(x|dx)$. This type of discrete transition is called \emph{%
spontaneous transition}. If the process reaches the boundary at $x\in 
\mathbf{X}_{\delta }$, the process has a discrete transition to a new random
state given by $R_{\delta }(x|dx)$. This type of discrete transition is
called \emph{forced transition}.

\noindent Always, we assume that $R_{o}(x,\mathbf{X}_{o})=1$ and $R_{\delta
}(x,\mathbf{X}_{o})=1$.

Thus, a sample trajectory has the form $(q_{t},x_{t},t\geq 0),$ where $%
(x_{t},t\geq 0)$ is piecewise continuous and $q_{t}\in Q$ is piecewise
constant. Let $0\leq T_{1}<T_{2}<...<T_{i}<T_{i+1}<...$ be the sequence of
jump times. The resulting process is a Borel right process as in the case of
PDMPs.

\section{Microscopic Level}

Let us consider an ULSoS of agents, whose behaviour exhibits discrete and
continuous dynamics with uncertainty features. Suppose that the ULSoS has a
large number of agents, each one having the dynamics described by a
diffusive-type model of stochastic hybrid system (that will be described
below). In this paper, we consider the case when the agent interactions will
change the\ hybrid structure of its behaviour by keeping the original
discrete transitions, but adding new discrete transitions as result of the
alteration of the continuous dynamics. More specifically, the continuous
dynamics for an agent mode can be modified using inputs coming from other
agents, i.e., it might encounter \emph{new discrete transitions} dictated by
these interactions. Then one operational mode is split in some new modes
resulted from the interaction between the hybrid agent and the entire
collective. To add more flexibility to the models of stochastic hybrid
systems used for the agent modelling, we consider that the mode boundaries
(guards) are not fixed in time. To achieve this, we need to allow guards
that exhibit dynamics governed by some ordinary differential equations (ODE).

\subsection{Hybrid Agent Model}

The mathematical model for a hybrid agent is a stochastic hybrid system with
a peculiar structure. The system has two types of discrete transitions:

\begin{itemize}
\item  event triggered transitions, which are generated by the detection of
certain events. In the context of the massively parallel collective, these
events are generated by the inter-agent communication, when an input message
is received.

\item  forced transitions, which are triggered by guards that can evolve in
time.
\end{itemize}

The system has input output activities, which follow a certain communication
policy. An output is sent only when a forced discrete transition takes
place. The inputs are collected only during continuous evolutions.

The continuous evolution of a hybrid agent has also a parallel structure by
executing simultaneously two distinct modes. Let us call these two modes as
the \emph{coordination}, respectively the \emph{activity }mode. These modes
start and stop synchronously. The coordination mode is, in fact, a hybrid
system by itself. Every input generates a discrete jump, and modifies the
dynamics in the coordination mode.

Formally, the activity of each agent is described as a stochastic hybrid
process $\mathbf{M}^{i}=(q_{t}^{i},z_{t}^{i},u_{t}^{i})$ (viewed as a sort
of revival process) defined on the hybrid state space $\mathbf{X}^{i}\times 
\mathbf{U}^{i}$, with some `active' (time depend) guards defined on $\mathbf{%
X}^{i}$. Note the definition of $\mathbf{X}^{i}$ should be slightly
different with respect to the classical case when the boundaries are fixed.
To avoid undesired complications, we suppose that for all $i$, the hybrid
state spaces $\mathbf{X}^{i}$ can be embedded in the Euclidean space $\Bbb{R}%
^{d}$. The pair $(q_{t}^{i},z_{t}^{i})$ will be called the coordination
component, and the pair $(q_{t}^{i},u_{t}^{i})$ will be called activity
component. The guards are defined as `active boundaries' or thresholds $%
(\partial _{t}^{i}):=(q_{t}^{i},\beta _{t}^{i})$ that replace the fixed
boundaries from the standard definition of SHS. More precisely, in the
absence of interactions, between the jump times, the process follows the
dynamics law given by some stochastic differential equations (\ref{SDE_mode}%
). The jumping times are defined as hitting times of the active boundaries.
Therefore, the evolution of such a hybrid system will be described by the
tuple $(q_{t}^{i},z_{t}^{i},\beta _{t}^{i}),$ $t\geq 0$, which is a right
continuous stochastic process on the underlying probability space $(\Omega ,%
\mathcal{F},\Bbb{P})$.

For each agent $i$, it is assumed that $z_{t}^{i}<\beta _{t}^{i}$ (the order
is defined componentwisely in the Euclidean space), for all $t\geq 0$,
except for the jumping moments of time $0<T_{1}^{i}<T_{2}^{i}<...$, when $%
z_{(T_{k}^{i}-)}^{i}=\beta _{(T_{k}^{i}-)}^{i}$. The active boundary is
thought of as a moving barrier $\{\beta _{t}^{i}|t\geq 0\}$, with $\beta
_{0}^{i}=\gamma ^{i}$. The jumping times are defined as the first hitting
times of the moving barrier by the continuous process $(z_{t}^{i})$. The
dynamics of the barrier will be given by a simple first order differential
equation: 
\begin{equation}
\frac{d\beta ^{i}}{dt}=\digamma ^{i}(\beta ^{i})\text{, }t>0\text{, }\beta
_{0}^{i}=\gamma ^{i}\text{.}  \label{barrier_dyn}
\end{equation}

\medskip The active boundary dynamics $\varphi ^{i}(t)$, which is a curve
defined respectively for each mode of the underlying agent $i$, is the
solution of (\ref{barrier_dyn}). Based on the hybrid nature of the
underlying system, we can think that the moving barrier is defined
piecewisely for each mode. Then, we can refine (\ref{barrier_dyn})
accordingly. In practice, we need to consider particular classes of ODE to
define the barrier dynamics, such that moments or probability distributions
of the jumping times can be analytically or numerically computed.
Considering the computation difficulty of the first time passage problem
when the active barrier has a quite general form (see \cite{Tuckwell}), we
specialize (\ref{barrier_dyn}) such that 
\begin{equation}
\frac{d\beta ^{i}}{dt}=(-k^{i})\cdot \beta ^{i}\text{, }t>0\text{; }\beta
_{0}^{i}=\gamma ^{i}\text{.}  \label{barrier_part}
\end{equation}
Then, the active boundaries have an exponential form $\beta _{t}^{i}=\gamma
^{i}e^{-k^{i}\Upsilon _{t}^{i}}$ such that the computation of the
expectations of the jumping times admits numerical solutions.

In the initial hybrid model of the agent $i$, we can also impose a reset
condition for the boundary variable $\partial =(q,\beta )$, as a stochastic
kernel 
\[
R_{\partial }^{i}:\Bbb{R}^{d}\times \mathcal{B}(\Bbb{R}^{d})\rightarrow [0,1]%
\text{.} 
\]
The role of $R_{\partial }^{i}$ is to provide the probability law for the
initial condition $\gamma ^{i}$ of the ODE that governs the guard dynamics.

Intuitively, an agent can have an independent evolution, or one which is
coordinated with the collective. When it evolves independently, the agent
executes an activity for a relatively long time until a forced transition
takes place and the agent switches to a different activity. For example, the
activities can be navigation under a certain direction, or rest. More
sophisticated scenarios could include the detection of a malign tumour, or
drug delivery. Different views can be used in defining activities. These can
be just simply labels from a given finite set, or they can be modelled into
more details by differential equations. One can add more details by
considering noise or other perturbations, and the mathematical model changes
into a stochastic differential equation. The collective behaviour requires a
formal coordination mechanism. This mechanism consists of perturbations of
the dynamics leading to a forced transition, generated by other agents via
communication. In order to keep the model simple, we make the assumption
that the inter-agent communication is not affecting directly an agent
activity. Instead, only the dynamics in the coordination mode is affected.
Since the guards of the forced transitions are related to the dynamics in
the coordination mode, the communications can speed up the execution of a
forced transition (and in this way make an activity change). A single
communication may not trigger a forced transition. Some times a repeated
communication from a single agent, or communications from several agents are
necessary. Equally, a communication can speed or slow the process of
executing a forced transition. These aspects are relevant for the problem of
stability, which is not treated in this paper.

The communication takes place along a bidirectional channel. Every hybrid
agent communicates only with a finite number of other agents called its
neighborhood. Each communication consists of a single bit. Practically, when
an agent executes a forced transition, all its neighbors are announced about
that.

Let formally model this sort of interaction between the individuals of a
ULSoS.

\medskip For each agent $i$, let define $\Upsilon _{t}^{i}:=t-T_{k}^{i}$ if $%
t\in [T_{k}^{i},T_{k+1}^{i})$, for $t\geq 0$. $\Upsilon _{t}^{i}$ denotes
the time elapsed since the last forced jump of the $i$th agent until the
moment $t$. Clearly, the discrete state is constant between jumps, i.e., $%
q_{t}^{i}=q_{T_{k}^{i}}^{i}$ if $t\in [T_{k}^{i},T_{k+1}^{i})$.

\begin{remark}
\label{Remark_clock}When $t\in [T_{k}^{i},T_{k+1}^{i})$, the time $\Upsilon
_{t}^{i}(\omega )$ can be thought of as a local clock for that part of the
trajectory $\omega $ that lies in the mode $q_{T_{k}^{i}}^{i}$. Then there
is a one-to-one correspondence between the trajectories of the hybrid agent $%
i$ and the trajectories of $(\Upsilon _{t}^{i})$.
\end{remark}

\noindent The dynamics of the agent $i$, for the coordination mode $%
q_{T_{k}^{i}}^{i}$, i.e., $t\in [T_{k}^{i},T_{k+1}^{i})$, \ is hybrid
discrete continuous. Within the interval between two consecutive
communication events, the dynamics is continuous. When a communication event
takes place, a discrete transition is produced, and then the continuous
dynamics changes. Let us formulate the analytics of this process. To each
unidirectional communication along the channel between the agents $i$ and $j$%
, we associate a characteristic vector $w^{ij}\in \Bbb{R}^{d}$.

\medskip For each agent $i$, let us consider the overall changes in the
dynamics due to all communications with its neighbors $I_{t}^{i}$ given by 
\[
I_{t}^{i}:=\sum_{j\in N^{i}}w^{ij}\exp (-k^{i}\Upsilon _{t}^{j}), 
\]
where $N^{i}$ is the neighborhood of the agent $i$, i.e. 
\[
N^{i}:=\{j:|w^{ij}|\geq w\text{; }\Upsilon _{t}^{j}\leq \Upsilon _{t}^{i}\}%
\text{,} 
\]
where $w>0$ is a lower threshold for the strength of interaction.

After all communications took place, the dynamics in the coordination mode
is given by the following equation 
\[
\widetilde{z}_{t}^{i}:=z_{\Upsilon _{t}^{i}}^{i}+I_{t}^{i}\text{, }t\in
[T_{k}^{i},T_{k+1}^{i})\text{.} 
\]

Now, we explain how the dynamics changes after each communication via a
recursive process where we define $(\widetilde{z}_{t}^{i})$ recurrently, as
follows. Suppose that 
\[
T_{k}^{i}\leq T_{k_{1}}^{j_{1}}<T_{k_{2}}^{j_{2}}<...<T_{k_{p}}^{j_{p}}\leq
t<T_{k+1}^{i}. 
\]
Then 
\begin{eqnarray*}
(\widetilde{z}_{t}^{i})_{1} &:&=z_{\Upsilon _{t}^{i}}^{i}+w^{ij_{1}}\exp
(-k^{i}\Upsilon _{t}^{j_{1}})\text{,} \\
(\widetilde{z}_{t}^{i})_{r} &:&=(\widetilde{z}_{t}^{i})_{r-1}+w^{ij_{r}}\exp
(-k^{i}\Upsilon _{t}^{j_{r}})\text{; }r=2,..,p\text{.}
\end{eqnarray*}
It is clear that each jump of the external agent $j$ (with respect to the
agent $i$) enables a jump in the continuous dynamics of agent $i$ of length $%
w^{ij}$. Roughly speaking, $I_{t}^{i}$ is forcing the apparition of some
discrete transition of the agent $i$ due to the communication with the
agents that have already exhibited such transitions.

\begin{figure}[h]
\begin{center}
\includegraphics[width=2.5in]{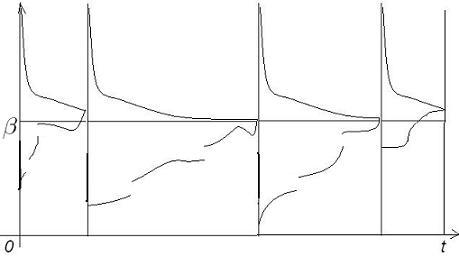}
\end{center}
\caption{Evolution of a hybrid agent}
\end{figure}

Figure 1 illustrates the evolutions of a one-dimensional hybrid agent. The
forced transitions are marked by a vertical line. The horizontal line marks
the asymptotic limit of the dynamic guard. A forced transition is triggered
when the guard and the dynamics in the coordination mode reach a common
value. One can easily remark the jumps determined by the communication
events.

\subsection{\protect\medskip Equivalent Description}

The first passage time of the active boundary corresponding to the
stochastic hybrid process $(\widetilde{z}_{t}^{i})$ can be thought also as
the first hitting time of a modified boundary corresponding to the initial
process $(z_{t}^{i})$. The new boundary continuous variable can be obtained
as: 
\begin{equation}
\overline{\beta }_{t}^{i}:=\beta _{t}^{i}-I_{t}^{i}.  \label{modified_guard}
\end{equation}

This means that the interactions between agents can be thought also as
acting on guards: The law of the active boundary of an agent is changed
according with the inputs coming from the other agents when they have a
jump. In other words, a change of the operational mode of one hybrid agent
in the ULSoS influences the guards (in our case, the active boundaries) of
the other agents. This new perspective on the type of interaction that we
have already defined here will help us for developing the analytical tools
for the macroscopic level of a ULSoS.

To capture the complexity of a one agent dynamics and its interactions with
other agents, we need to consider also the time process $\Upsilon _{t}^{i}$,
i.e. 
\begin{equation}
(a_{t}^{i}):=(q_{t}^{i},\widetilde{z}_{t}^{i},\beta _{t}^{i},\Upsilon
_{t}^{i}),\text{ }t\geq 0.  \label{hybrid_agent}
\end{equation}
Due to the interaction with the other agents, $(a_{t}^{i})$ is not
necessarily a Markov process. However, it becomes a Markov process if there
are no interactions, or if we fix $\Upsilon _{t}^{j}$ for the agents $j\in
N^{i}$. Moreover, the evolution of the process $(q_{t}^{i},\widetilde{z}%
_{t}^{i})$ can be encoded in the evolution of $(q_{t}^{i},\beta
_{t}^{i},\Upsilon _{t}^{i})$. Then the agent activity will be driven only by
the boundary dynamics and the jumping times encapsulated in $\Upsilon
_{t}^{i}$, i.e., $(\partial _{t}^{i},\Upsilon _{t}^{i})$. A similar idea has
been used in the use of jump processes for studying Piecewise Deterministic
Markov Processes (PDMP) - see \cite{DA93}. Note that in our case $(\partial
_{t}^{i},\Upsilon _{t}^{i})$ is not a jump process, but it is a hybrid
process with deterministic continuous dynamics.

Let $\varsigma ^{i}$ be the first hitting time of $(z_{t}^{i})$ to reach the
curve $(\overline{\beta }_{t}^{i})$ defined by (\ref{modified_guard}). Let
us collect all

\begin{itemize}
\item  $N$ clock variables as $\tau :=(\tau ^{1},...,\tau ^{N})$,

\item  $N$ guard variable as $\beta :=(\beta ^{1},...,\beta ^{N})$ (note
that $\beta $ is an $N\times d$ dimensional vector).
\end{itemize}

Define $\tau ^{(-i)}:=(\tau ^{1},...,\tau ^{i-1},\tau ^{i+1},...,\tau ^{N})$%
. Let $I$ be a measurable set of $[0,\infty ),$ and let us define (as a
conditional probability) the following measure: 
\[
\mu _{\beta }^{i}(I):=\Bbb{P}[\varsigma ^{i}\in I|\tau ^{(-i)},\beta ]\text{.%
}
\]
Suppose that there is a probability density function associated to $\mu
_{\beta }^{i}$, i.e., $\psi ^{i}(t|\tau ^{(-i)},\beta )dt:=\mu _{\beta
}^{i}(dt)$ and define $\Psi ^{i}$ as the probability that the stopping time $%
\varsigma ^{i}$ is less than $\tau ^{i}$, i.e., 
\[
\Psi ^{i}(\tau ^{i}|\tau ^{(-i)},\beta ):=\mu _{\beta }^{i}([0,\tau
^{i}])=\int_{0}^{\tau ^{i}}\psi ^{i}(t|\tau ^{(-i)},\beta )dt
\]

\begin{theorem}
$(\beta _{t}^{i},\Upsilon _{t}^{i})$ is a Piecewise Deterministic Markov
Process.
\end{theorem}

%TCIMACRO{\TeXButton{Proof}{\proof}}
%BeginExpansion
\proof%
%EndExpansion
The standards features that characterize a PDMP are: deterministic dynamics
for the continuous evolution, discrete transitions (governed either by a
rate function, or by guards), and a reset map defined as stochastic kernel.
In our case, it is clear that the continuous dynamics is governed by simple
ODEs. The process $\Upsilon _{t}^{i}$ is just simply increasing with the
unit rate, and has reset to zero whenever $z_{t}^{i}$ has a jump. Moreover,
our process $(\beta _{t}^{i},\Upsilon _{t}^{i})$ does not have forced jumps
(due to the existence of guards). The discrete transitions take place in a
Poisson type fashion with respect to a rate function. We identify this rate
function as the following measurable bounded function 
\begin{equation}
\lambda ^{i}(\tau ,\beta ):=\frac{\psi ^{i}(\tau ^{i}|\tau ^{(-i)},\beta )}{%
1-\Psi ^{i}(\tau ^{i}|\tau ^{(-i)},\beta )}\text{.}  \label{jump_rate}
\end{equation}
This has the role of a transition rate: the probability that the agent $i$
has a jump in the interval $\Delta t$ is equal to $\lambda ^{i}(\tau ,\beta
)\Delta t+o(\Delta t)$. The probability that in the interval $\Delta t$, two
or more agents may have discrete transitions is $o(\Delta t)$. The reset
kernel is trivial $(R_{\partial }^{i},R_{0}^{i})$, where $R_{\partial }^{i}$
is the reset kernel for the guard, and $R_{0}^{i}$ is the reset to $0$ of
the clock variable. 
%TCIMACRO{\TeXButton{End Proof}{\endproof}}
%BeginExpansion
\endproof%
%EndExpansion

\begin{remark}
$(\beta _{t}^{i},\Upsilon _{t}^{i})$ is a PDMP with spontaneous discrete
transitions governed by $\lambda ^{i}(\tau ,\beta )$, but no forced discrete
transitions.
\end{remark}

\section{Macroscopic Level}

\smallskip The ULSoS collective behaviour is described by the interactive
superposition of its hybrid agents $(q_{t}^{i},z_{t}^{i},\beta
_{t}^{i},\Upsilon _{t}^{i})$, $i=1,..N$. We denote this superposition as
follows 
\[
\otimes ^{\blacklozenge }{}(q_{t}^{i},z_{t}^{i},\beta _{t}^{i},\Upsilon
_{t}^{i}) 
\]
The entire ULSoS activity is completely described by the embedded Markov
hybrid process $(\partial _{t},\Upsilon _{t})$ defined on the hybrid state
space obtained as the superposition of the agent state spaces $\bigcup_{q\in
Q^{i}}\{q\}\times \Bbb{R}^{d}\times [0,\infty )$. In fact, the boundary
variable $\beta $ is carrying in the structure also information about the
discrete state $q$. Therefore, to simplify the up-coming analysis, we need
only the process $(\beta _{t},\Upsilon _{t})$ defined on $\Bbb{R}^{d\times
N}\times [0,\infty )^{N}$. Such a process play the role of a macroscopic
`abstraction' for the ULSoS behaviour. The executions of the process $(\beta
_{t},\Upsilon _{t})$ can be described as follows. Each component $\beta
_{t}^{i}$ of $\beta _{t}$ follows the dynamics described by (\ref
{barrier_part}) in the interval between two forced transitions of the agent $%
i$, whereas each component $\Upsilon _{t}^{i}$ of $\Upsilon _{t}$ follows a
trivial ODE with the rate $1.$ As an easy consequence of the Remark \ref
{Remark_clock}, we get the following result.

\begin{theorem}
For any initial condition, there is a one-to-one correspondence between the
sample paths of \noindent $\otimes ^{\blacklozenge
}{}(q_{t}^{i},z_{t}^{i},\beta _{t}^{i},\Upsilon _{t}^{i})$ and $(\beta
_{t},\Upsilon _{t}).$
\end{theorem}

%TCIMACRO{\TeXButton{Proof}{\proof}}
%BeginExpansion
\proof%
%EndExpansion
The proof can be done in the same style used by Davis \cite{DA93} to prove
that there is a one-to-one correspondence between the paths of a jump
process and the paths of a PDMP.%
%TCIMACRO{\TeXButton{End Proof}{\endproof}}
%BeginExpansion
\endproof%
%EndExpansion

The process $(\beta _{t},\Upsilon _{t})$ will be called \emph{ULSoS
abstraction}. We need to clarify this concept, because all the further
developments of this work are based on this. Recall that at the microscopic
level we have modelled the agents hybrid behaviour and their interactions
using a sort of communicating stochastic hybrid systems. Then the collective
behaviour of the ULSoS is a complex stochastic hybrid system obtained by
sticking together the agent dynamics described by (\ref{hybrid_agent}). The
analysis of this new hybrid system requires indeed complicated mathematics,
since we have to consider different facets: stochastic differential
equations, discrete transitions governed by dynamic guards, interaction
between agents that might change the continuous dynamics, and so on. The
observation that from all parameters that describe the ULSoS behaviour, only
two of them can be used to construct a sort of skeleton process that
characterizes the entire ULSoS dynamics is essential for this analysis. This
point is crucial for developing our approach for finding useful
characterizations of the ULSoS at the macroscopic level. Moreover, we can go
further with the abstraction process and to derive form the given PDMP the
embedded Markov jump process.

For a given collective, one can designate some hybrid agents to play the
role of input, and similarly some output agents. In this way, the
collectives can be composed by connecting one collective's output to the
input of the second. At the macroscopic level, the composite collective is
described by sequential composition of PDMPs. Various composition operators
for PDMPs form the so-called process algebra. This has been developed in 
\cite{BBM05}. This is the key for the modular development of multi-agent
collectives.

For the macroscopic description of the ULSoS, it is necessary to provide
some PDE to describe the dynamics of the spatially distributed agent
densities. In our case, the abstraction process $(\beta _{t},\Upsilon _{t})$
is embedded in the dynamics structure of the entire ULSoS. This abstraction
process is bidirectional: some properties of the abstraction process will
characterize also the whole ULSoS dynamics, but also the agent dynamics. For
the ULSoS modelling framework we have defined here, the PDE that will arise
naturally as backward Kolmogorov equation, or forward Kolmogorov
(Fokker-Planck) equation associated to the ULSoS abstraction. Such equations
will describe the evolution of the guards and local clocks probability
distributions. The first step for the derivation of such equations is to
obtain the mathematical expression of the \emph{infinitesimal generator}
associated to $(\beta _{t},\Upsilon _{t})$.

\subsection{Infinitesimal Generator}

Let us briefly recall the concept of infinitesimal generator. Suppose that $%
(x_{t})$ is a Markov process with an homogeneous transition probability
function $(p)_{t\geq 0}$. For each $t\geq 0$, define conditional expectation
operator by 
\begin{equation}
\mathbf{P}_{t}f(x):=\int f(y)p_{t}(x,dy)=\Bbb{E}_{x}f(x_{t}),\forall x\in 
\mathbf{X}\text{;}  \label{semigr}
\end{equation}
where $\Bbb{E}_{x}$ is the expectation with respect to $\Bbb{P}_{x}$. Here, $%
f$ belongs to $\mathcal{B}_{b}(\mathbf{X})$, which is the lattice of all
bounded measurable real functions defined on $\mathbf{X}$. The
Chapman-Kolmogorov equation guarantees that the linear operators $\mathbf{P}%
_{t}$ satisfy the semigroup property: $\mathbf{P}_{t+s}=\mathbf{P}_{t}%
\mathbf{P}_{s}$. This suggests that the semigroup of (conditional
expectation) operators $\mathcal{P}=(\mathbf{P}_{t})_{t>0}$ can be
considered as a sort of parameterization for a Markov process.

Associated with the semigroup $(\mathbf{P}_{t})$ is its \emph{infinitesimal
generator} which, loosely speaking, is the derivative of $\mathbf{P}_{t}$ at 
$t=0$. Let $D(L)\subset \mathcal{B}^{b}(\mathbf{X})$ be the set of functions 
$f$ for which the following limit exists 
\begin{equation}
\lim_{t\searrow 0}\frac{1}{t}(\mathbf{P}_{t}f-f)  \label{generator}
\end{equation}
and denote this limit $Lf$. The limit refers to convergence in the supnorm $%
\left\| \cdot \right\| $ of the Banach space $\mathcal{B}_{b}(\mathbf{X})$,
i.e. for $f\in D(L)$ we have: 
\[
\lim_{t\searrow 0}||\frac{1}{t}(\mathbf{P}_{t}f-f)-Lf||=0. 
\]

For a PDMP defined as in Section \ref{sect_PDMP}, the infinitesimal
generator has the following expression 
\begin{equation}
\mathcal{L}f(x)=b(x)\cdot \nabla f(x)+\lambda (x)\int [f(y)-f(x)R(x,dy)
\label{gen_PDMP}
\end{equation}
for any $f\in D(\mathcal{L})$. The domain of the generator $D(\mathcal{L})$
is fully described in \cite{DA93}.

Now, coming back to our process $(\beta _{t},\Upsilon _{t})$, it is clear,
from the construction, that this new process is obtained by the interacting
PDMP components. Then, the expression of its infinitesimal generator will be
based on the well known generator expression for PDMP.

For a better understanding, we derive first the expression of the
infinitesimal generator corresponding to a hybrid agent viewed as a PDMP.

\begin{proposition}
The infinitesimal generator associated to $(\beta _{t}^{i},\Upsilon
_{t}^{i}) $ maps a continuous differentiable function $f^{i}(\beta ^{i},\tau
^{i})$: $\Bbb{R}^{d}\times [0,\infty )\rightarrow \Bbb{R}$ as follows: 
\begin{equation}
\begin{tabular}{ll}
$L^{i}f^{i}(\beta ^{i},\tau ^{i})$ & $=\lambda ^{i}(\beta ^{i},\tau
^{i})\int_{\Bbb{R}^{d}}(f^{i}(\theta ^{i},0)-f^{i}(\beta ^{i},\tau
^{i}))R_{\partial }^{i}(\beta ^{i},d\theta ^{i})$ \\ 
& $+(-k^{i})\sum_{p=1}^{d}\beta _{p}^{i}\frac{\partial f}{\partial \beta
_{p}^{i}}(\beta ^{i},\tau ^{i})+\frac{\partial f}{\partial \tau ^{i}}(\beta
^{i},\tau ^{i}).$%
\end{tabular}
\label{gen_agent}
\end{equation}
\end{proposition}

%TCIMACRO{\TeXButton{Proof}{\proof}}
%BeginExpansion
\proof%
%EndExpansion
Applying directly the general formula (\ref{gen_PDMP}), with the reset
kernel $R_{\partial }^{i}\otimes R_{0}^{i}$. The effect of $R_{0}^{i}$ is
the apparition of $0$ in $f^{i}(\theta ^{i},0)$ in the integral part of (\ref
{gen_agent}).%
%TCIMACRO{\TeXButton{End Proof}{\endproof}}
%BeginExpansion
\endproof%
%EndExpansion

\begin{remark}
\medskip If one would consider to work with more general dynamics (\ref
{barrier_dyn}), then the differential part of the generator expression (\ref
{gen_agent}) has to be changed accordingly with the more general formula (%
\ref{gen_PDMP}).
\end{remark}

Let us define the vector field $b:\Bbb{R}^{N\times (d+1)}\rightarrow \Bbb{R}%
^{N\times (d+1)}$ that describe the continuous evolution of $(\beta
_{t},\Upsilon _{t})$ as follows:\noindent 
\[
b:=(-k^{1}\beta _{1}^{1},...,-k^{1}\beta _{d}^{1},...,-k^{N}\beta
_{1}^{N},...,-k^{N}\beta _{1}^{N},1,1,...,1). 
\]

Given a function $f\in \mathcal{C}^{1}(\Bbb{R}^{N\times (d+1)},\Bbb{R})$ and
a vector field $b$, we use $\mathcal{L}_{b}f$ to denote the \emph{Lie
derivative} of $f$ along $b$ given by 
\begin{eqnarray*}
\mathcal{L}_{b}f(\beta ,\tau ) &=&\sum_{p=1}^{N\times (d+1)}\frac{\partial f%
}{\partial (\beta ,\tau )_{p}}(\beta ,\tau )b_{p}(\beta ,\tau ) \\
&=&\sum_{i=1}^{N}(-k^{i})\sum_{p=1}^{d}\beta _{p}^{i}\frac{\partial f}{%
\partial \beta _{p}^{i}}(\beta ,\tau )+\sum_{i=1}^{N}\frac{\partial f}{%
\partial \tau ^{i}}(\beta ,\tau )
\end{eqnarray*}
where 
\begin{eqnarray}
(\beta ,\tau ) &:&=(\beta ^{1},\beta ^{2},..,\beta ^{N},\tau )\text{;}
\label{state_space_vector} \\
\beta ^{1} &:&=(\beta _{1}^{1},...,\beta _{d}^{1})\text{; }\beta
^{2}:=(\beta _{1}^{2},...,\beta _{d}^{2})  \nonumber \\
&&...  \nonumber \\
\beta ^{N} &:&=(\beta _{1}^{N},...,\beta _{d}^{N})\text{; }\tau :=(\tau
^{1},...,\tau ^{N}).  \nonumber
\end{eqnarray}

Let us define a stochastic kernel $R_{\partial }$ obtained by the
superposition of the corresponding kernels for all agents, i.e. 
\begin{eqnarray*}
R_{\partial } &:&\Bbb{R}^{N\times d}\times \mathcal{B}(\Bbb{R}^{N\times
d})\rightarrow [0,1] \\
R_{\partial } &:&=R_{\partial }^{1}\otimes R_{\partial }^{2}\otimes
...R_{\partial }^{N}\text{.}
\end{eqnarray*}
Now we have all the elements to write down the infinitesimal generator of $%
(\beta _{t},\Upsilon _{t})$.

We use the notation $\beta ^{i}(\theta )$ to express the fact that, in the
expression (\ref{state_space_vector}), the component $\beta ^{i}$ has been
replaced by $\theta $, and the notation $\tau ^{i}(0)$ to say that, in the
same expression, the element $\tau ^{i}$ has been replaced by $0$.

\begin{theorem}
The infinitesimal generator associated to $(\beta _{t},\Upsilon _{t})$ can
be expressed as follows: 
\begin{equation}
Lf(\beta ,\tau )=L_{cont}f(\beta ,\tau )+L_{jump}f(\beta ,\tau )
\label{gen_swarm}
\end{equation}
where 
\[
L_{cont}f(\beta ,\tau ):=\mathcal{L}_{b}f(\beta ,\tau ), 
\]
and

\[
L_{jump}f(\beta ,\tau ):=\sum_{i=1}^{N}\left\{ \lambda ^{i}(\beta ,\tau
)\noindent \cdot \int_{\Bbb{R}^{d\times N}}(f(\beta ^{i}(\theta ),\tau
^{i}(0))-f(\beta ,\tau ))R_{\partial }^{i}(\beta ^{i},d\theta )\right\} . 
\]
\end{theorem}

%TCIMACRO{\TeXButton{Proof}{\proof}}
%BeginExpansion
\proof%
%EndExpansion
$(\beta _{t},\Upsilon _{t})$ is a PDMP obtained by the interaction of the
PDMP components. Then the expression of the infinitesimal generator follows
the general expression of a PDMP generator, taking also into account the
interacting factors.%
%TCIMACRO{\TeXButton{End Proof}{\endproof}}
%BeginExpansion
\endproof%
%EndExpansion

\begin{remark}
The infinitesimal generator of the ULSoS abstraction is obtained by summing
the generators of the PDMP components. The interaction between the
components is captured only by the transition rates $\lambda ^{i}(\beta
,\tau )$.
\end{remark}

\subsection{PDE Characterizations}

Departing from the expression of the infinitesimal generator of the ULSoS
abstraction, one can obtain the PDE associated. In the following, we give a
short background on the Kolmogorov equations associated to a Markov process,
and then explain the peculiarities of such equations for stochastic hybrid
systems, and, in particular, for ULSoS.

\subsubsection{Kolmogorov Equations for Markov Processes}

This subsection recalls some basic facts concerning the backward and forward
Kolmogorov equation for Markov processes. The forward equation is also known
as the Fokker Planck Kolmogorov (FPK) equation for diffusion processes. The
Fokker Planck equation is one of the basic tools when dealing with diffusion
processes, because it allows to calculate the probability density function
(pdf) $\rho _{t}$ of the process at time $t\geq 0$ given an initial
probability density $\rho _{0}$ and eventually the stationary pdfs (when
they exist).

The semigroup $(\mathbf{P}_{t})$ of a Markov process $\mathbf{M}=(x_{t})$
satisfies the following differential equation: for all $f\in D(L)$, 
\begin{equation}
\frac{d}{dt}\mathbf{P}_{t}f=L\mathbf{P}_{t}f\text{.}  \label{backward_K}
\end{equation}
This equation is called \emph{Kolmogorov's backward equation} \cite{DA93}.
In particular, if we define the function $u(t,x)=\mathbf{P}_{t}f(x)$ then $u$
is solution of the PDE

\begin{center}
$\left\{ 
\begin{tabular}{l}
$\frac{\partial u}{\partial t}=Lu$ \\ 
$u(0,x)=f(x).$%
\end{tabular}
\right. $
\end{center}

\noindent Conversely, if this PDE admits a unique solution, then its
solution is given by $\mathbf{P}_{t}f(x)$. Moreover, it is easy to check
that the operators $\mathbf{P}_{t}$ and $L$ commute. Then (\ref{backward_K})
may be written as 
\begin{equation}
\frac{d}{dt}\mathbf{P}_{t}f=\mathbf{P}_{t}Lf.  \label{forward_k}
\end{equation}
This equation is known as \emph{Kolmogorov's forward equation}. It is the
weak formulation of the equation $\frac{d}{dt}\mu _{t}^{x}=L^{*}\mu _{t}^{x}$%
, where the probability measure $\mu _{t}^{x}$ on $\mathbf{X}$ denotes the
law of $(x_{t})$ conditioned on $x_{0}=x$ and where $L^{*}$ is the adjoint
operator of $L.$

In particular, if $\mathbf{M}$ is a diffusion process on $\Bbb{R}^{n}$ and
if $\mu _{t}^{x}(dy)$ admits a density $q(x;t,y)$ with respect to. the
Lebesgue measure, the forward Kolmogorov equation is the weak form (in the
sense of distribution theory) of the PDE 
\begin{equation}
\frac{\partial }{\partial t}q(x;t,y)=-\sum_{i=1}^{n}\frac{\partial }{%
\partial y_{i}}(b_{i}(y)q(x;t,y))+\sum_{i,j=1}^{d}\frac{\partial ^{2}}{%
\partial y_{i}\partial y_{j}}(w_{ij}(y)q(x;t,y))\text{,}  \nonumber
\end{equation}
where $b_{i}(x)$ and $w_{ij}(x)$ are respectively the drift coefficient and
the diffusion coefficient of the process. This equation is known as the
Fokker-Planck equation associated to a diffusion process.

\subsubsection{Kolmogorov Equations for ULSoS}

The macroscopic description of a ULSoS is described by a Markov jump type
process. Here, jump process is understood in a rather large sense, i.e.
process with discontinuities in the natural filtration. A complete
description of a Markov jump process is given by its transition density
function, which is the solution of the forward and backward Kolmogorov
equations.

A generalized Fokker Planck equation is well known for the case of switching
diffusions (where there are no forced transitions). \ A unifying formulation
of the Fokker-Planck-Kolmogorov equation for general stochastic hybrid
systems is developed in \cite{Bect2008}. For some particular PDMPs, FPK
equation has proved to be an useful tool for studying multi-agent systems 
\cite{Mesquita}.

The FPK equation for stochastic hybrid systems\ is based on the concept of
mean jump intensity. Let us define a positive measure $J$ on $\mathbf{X}%
\times (0,\infty )$ by 
\[
J(A)=\Bbb{E}_{\mu _{0}}\{\sum_{k\geq 0}1_{A}(x_{T_{k}}^{-},T_{k})\}. 
\]
For any $\Gamma \in \mathcal{B}$, the quantity $J(\Gamma \times (0,t])$ is
the expected number of jumps starting from $\Gamma $ during the interval $%
(0,t]$.

\medskip Suppose that there exists a mapping $r:t\mapsto r_{t}$, from $%
[0,\infty )$ to the set of all bounded measures on $\mathbf{X}$ such that
for all $\Gamma \in \mathcal{B}$, we have: (a) $t\mapsto r_{t}(\Gamma )$ is
measurable; (b) for all $t\geq 0$, 
\[
J(\Gamma \times (0,t])=\int_{0}^{t}r_{l}(\Gamma )dl. 
\]
Then $r$ is called the \emph{mean jump intensity} of the process $\mathbf{M}$
under the initial law $\mu _{0}$.

The generalized FPK\ equation can be written symmetrically as 
\begin{equation}
\mu _{t}^{^{\prime }}=\mathcal{L}_{cont}^{*}\mu _{t}+\int (W_{t}(dx,\cdot
)-W_{t}(\cdot ,dx))  \label{FPK_SHS}
\end{equation}
where $W_{t}(dx,dy)=r_{t}(dx)R(x,dy)$ ($R$ is the stochastic kernel that
provides the probability distributions of the post jump locations), or 
\begin{equation}
\mu _{t}^{^{\prime }}=\mathcal{L}_{cont}^{*}\mu _{t}+r_{t}(R-I)
\label{FPK_simplu}
\end{equation}
where $I$ is the identity kernel, i.e. $I(x,dy)=\delta _{x}(dy)$. In (\ref
{FPK_SHS}), $\mu _{t}$ is the law of the process $x_{t}$, and $t\rightarrow
\mu _{t}^{^{\prime }}$ the derivative of $t\rightarrow \mu _{t}$ (in the
sense of measure theory). Here, $\mathcal{L}_{cont}^{*}$ is the adjoint of $%
\mathcal{L}_{cont}$ (the continuous part of the infinitesimal operator of $%
\mathbf{M}$) in the sense of distribution theory.

Remark that in the case of stochastic hybrid processes, the forward and
backward Kolmogorov equations are \emph{parabolic integro partial
differential equations}.

The backward/forward Kolmogorov equations of a stochastic hybrid process
that describes the ULSoS abstraction are based on the expression of the
infinitesimal generator (\ref{gen_swarm}).

\begin{remark}
\cite{Bect2008} For spontaneous jumps, a mean jump intensity always exists,
and it is the expectation of the transition rate function (stochastic jump
intensity) $\lambda (x_{t})$ on the event $\{x_{t}\in \Gamma \}$.
\end{remark}

This is a key remark for our analysis. Then the derivation of the FPK
equation seems to be feasible for the ULSoS abstraction process (since it
does not exhibit forced transitions). The only problem we encounter is that
the expression of the transition rate function (\ref{jump_rate}) for $(\beta
_{t}^{i},\Upsilon _{t}^{i})$ is not known! This rate depends on the
probability distribution function of the first passage time of the modified
active boundary. In the next section, we will exploit additional hypotheses
that can make the computation of these jump rates feasible.

\section{Conclusions}

In this paper, we have proposed a rigorous, mathematical modelling framework
for massively parallel multi-agent systems. The purpose of this framework is
to allow the top down control. Each agent is a stochastic hybrid system, and
the multi-agent system itself is also a stochastic hybrid system. The model
has two scales. At the microscopic scale, agents have a coordination
dynamics, and an activity dynamics. Each agent can communicate with its
neighborhood, a finite set of agents connected via bidirectional
communication channels. At the macroscopic level, the dynamics of an ULSoS
is modelled as a PDMP. The macroscopic level is useful for composing MPMASs.

The technical contribution consists of determining the expression of the
infinitesimal generator of an ULSoS, and the derivation of the associated
Kolmogorov equations.

In the future work, we will investigate the topics of logics for specifying
and reasoning about massive parallelism, probabilistic model checking of
safety and performability properties and identifying multiple scales that
can be related by formal refinement/abstraction relations.

To the author knowledge, the approach presented in this paper is new and
original. A continuous time continuous space model for swarms has been
developed in \cite{swarm}. Like in this work, there the author considers
also a two layers model. There are two major differences compared to this
approach. At macroscopic level, the model from \cite{swarm} is continuous
with no possibility of operating regime change. This makes the control more
difficult. The second difference comes from the control perspective. In the
above reference, the control is bottom-up studying the impact at the
macroscopic level of the simple interaction rules from the microscopic
level. The approach developed in \cite{Berman} (and the references therein)
is also based on hybrid systems. There, each hybrid agent is deterministic,
but the agents interact following the pattern of a chemical reaction
network, which is probabilistic. The control is also top-down and two
layered. At the macroscopic level, optimization strategies are investigated,
while at the microscopic level, the focus is on collision avoidance.

%\nocite{*}
\bibliographystyle{eptcs}
\bibliography{references}

\end{document}